\documentclass[runningheads]{llncs}
\usepackage[nolist]{acronym}

\usepackage{cite}
\usepackage{colortbl}
\usepackage{hyperref}
\usepackage{graphicx}
\usepackage{enumitem}

\usepackage{array}
\usepackage{diagbox}
\usepackage{colortbl}
\usepackage{hhline}
\usepackage{graphicx}
\usepackage{subcaption}
\usepackage[breakable, hooks]{tcolorbox}
\usepackage{tikz}

\begin{document}
\begin{acronym}
 \acro{OSI}{Open Systems Interconnection model}
 \acro{CPU}{Central Processing Unit}

 \acro{OTF}{OpenType Font}
 \acro{WOFF}{Web Open Font Format Version 1.0}
 \acro{WOFF2}{Web Open Font Format Version 2.0}
 \acro{TTF}{TrueType Font}
 \acro{WLAN}{Wireless Local Area Network}
 \acro{CSS}{Cascading Style Sheet}
\end{acronym}
\title{On the Role of Font Formats in Building Efficient Web Applications}

\titlerunning{Comparison of Web Font Formats}

\author{Benedikt Dornauer\inst{1,2}\orcidID{0000-0002-7713-4686} \and
Wolfgang Vigl \inst{1}\orcidID{0009-0007-4938-2782} \and
Michael Felderer\inst{1,2,3}\orcidID{0000-0003-3818-4442}}
\authorrunning{Dornauer et al.}

\institute{University of Innsbruck, 6020 Innsbruck, Austria \\ \email{\{benedikt.dornauer,wolfgang.vigl\}@uibk.ac.at}\and
University of Cologne, 50923 Cologne, Germany \and
German Aerospace Center (DLR), Institute for Software Technology, 51147 Cologne, Germany\\ 
\email{michael.felderer@dlr.de}}
\maketitle              
\begin{abstract}
The success of a web application is closely linked to its performance, which positively impacts user satisfaction and contributes to energy-saving efforts. Among the various optimization techniques, one specific subject focuses on improving the utilization of web fonts. This study investigates the impact of different font formats on client-side resource consumption, such as CPU, memory, load time, and energy. In a controlled experiment, we evaluate performance metrics using the four font formats: OTF, TTF, WOFF, and WOFF2. The results of the study show that there are significant differences between all pair-wise format comparisons regarding all performance metrics. Overall, WOFF2 performs best, except in terms of memory allocation. Through the study and examination of literature, this research contributes (1) an overview of methodologies to enhance web performance through font utilization, (2) a specific exploration of the four prevalent font formats in an experimental setup, and (3) practical recommendations for scientific professionals and practitioners.
\end{abstract}

\section{Introduction}
\label{sec:introduction}
To ensure the success of websites and achieve optimal user satisfaction, it is crucial to consider usability and various other design criteria \cite{nielsen2000,pearrow2000}. According to Took \cite{took1990}, users' interaction with websites with a higher-than-normal user experience was significantly associated with improved web performance. The importance of performance was also stated by Google's March 2016 data showing that more than half of mobile site visits are abandoned if the site takes longer than three seconds to load \cite{an2018}. Besides that, improving performance also contributes to energy-saving efforts, reducing the time and resources required for rendering and displaying content \cite{rodriguez2015}. Consequently, comprehensive research is in progress to enhance the efficiency of web applications with different approaches, including specific considerations related to the utilization of fonts. 

In an article from 2022, Google experts Hempenius and Pollard \cite{webdev} explore the potential performance enhancements of different web page font settings. They identify potential bottlenecks for web performance and provide insights into mitigation opportunities. One of their mentioned bottlenecks was the improper choice of suitable font formats. 

Usually, fonts are utilized in computer systems and other text presentation systems to represent glyphs visually. Most computers have various fonts available preinstalled for creating documents and graphics. This availability and flexibility of fonts are the culmination of over three decades of gradual advancements in computer font science \cite{wright}. Nowadays, fonts also play a pivotal role in impactful web designs, as font features convey feelings and reactions that text alone cannot achieve. Hence, fonts are often chosen that match the corporate design \cite{ tidwellDesigningInterfacesPatterns2020}.

Unfortunately, these specific font styles are often not preinstalled on the devices by default \cite{zhao2018}. In order to solve this requirement (using a non-standard font), web developers use a \texttt{@font-face} declaration in \ac{CSS} file to declare the new fonts. The declaration also includes an URL to the online font-file resource (e.g., Google Fonts), a file holding information about the specific custom font \cite{olsson2019}. 

Among the various font file formats, the predominant options encompass the system font formats, TrueType Font (TTF) and OpenType Font (OTF), as well as the web font formats, Web Open Font Format (WOFF) and its second generation WOFF2. Most of web browsers widely support TTF, OTF, and WOFF, whereas WOFF2 is comparatively supported only by newer browser versions. TTF and OTF hold particular significance due to their extensive availability, serving as standard font formats developed by Adobe, Microsoft, and Apple. While TTF and OTF are font formats designed for system fonts, WOFF and the newer version WOFF2 are web fonts optimized for loading from a web server. WOFF is a container format that embeds TrueType or OpenType fonts and compresses them. The second version of WOFF  can show a significantly reduced file size, according to Buhler et al. \cite{buhler2017}. Following the transmission of web fonts to the client, the browser undertakes data decompression to facilitate font loading and display. Although the diminished file size contributes to decreased transfer time, this advantage is anticipated to be accompanied by heightened CPU and memory utilization \cite{woffw3c, ouyang2010}.

This study investigates how the choice of font format in web applications affects the client´s device performance. Therefore, we conducted a benchmark experiment to investigate the effects during the loading of web content of \textbf{different font formats (independent variable)} on the \textbf{clients' performance-related metrics (dependent variable)}. Therefore, the research question is:\\
\begin{center}
   \textit{\textbf{[RQ] How do different web font formats compare in terms of their impact on performance improvement in web applications?}}
\end{center}
\newpage
In order to evaluate the performance-related metrics, the following null-hypothesis are defined: 
\begin{itemize}[leftmargin=2.0em] 
    \item[\textbf{$H_1$}]\label{hypo:h1} The font format does not influence the required \textbf{Document Loading Time}.
    \item[\textbf{$H_2$}]\label{hypo:h2} The font format does not affect the \textbf{Processor Cycles}.
    \item[\textbf{$H_3$}]\label{hypo:h3} The font format does not impact the \textbf{Allocated Amount of Memory}.
    \item[\textbf{$H_4$}]\label{hypo:h4} The font format overall does not influences the \textbf{Energy Consumption}.
\end{itemize}

After providing the necessary context and motivation for this experiment, the subsequent sections of this paper are organized as follows. Section \ref{sec:relatedWork} provides background information on font optimization techniques and related work, including grey literature. Section \ref{sec:experimentalMethodology} describes the conducted methodology. Subsequently, in Section \ref{sec:results}, the implications
of these findings are then discussed in \ref{sec:discussion} and followed by Section \ref{sec:threatsToValididty}, which addresses study limitations and potential Threats to Validity. Finally, Section \ref{sec:conclusion} concludes the overall study.

\section{Related Work}
\label{sec:relatedWork}
Prior studies on web font optimization have followed two distinct methods. The first perspective focuses on identifying improved visual representations that enhance task performance, promote better text comprehension, and ultimately increase user satisfaction ( \textit{\ref{sec:aestheticOptimization} Aesthetic Optimization}). The second viewpoint revolves around optimizing computational performance, such as by reducing the load time, which is the perspective that is considered mainly in this paper (\textit{\ref{sec:performanceOptimization} Performance Optimization}). 

\subsection{Aesthetic Optimization}
\label{sec:aestheticOptimization}
Ling et al. \cite{ling2006} have presented the results of two experiments in which the influence of font and line length on several task performance and subjective measures was investigated. The authors showed that the effect of line length was significant on performance, but the effect of font type had an insignificant small impact.

Similarly, Bhatia et al. \cite{bhatiaEffectFontSize2011} have conducted a study to investigate the effects of font size, italics, and color count on three web usability dimensions: effectiveness, efficiency, and satisfaction. While the effectiveness and efficiency of the participants were measured via tasks, satisfaction was determined using a survey instrument. The study showed that font size and number of colors had no significant effect on any variable. However, using italics had a statistically significant effect on performance but not on efficiency and satisfaction. 

In 2016, Rello et al. \cite{rello2016} examined the font size and line spacing in more detail regarding objective and subjective legibility and showed a continuous improvement in both up to a size of 18 points. From 22pt, there was again a decrease in subjective legibility. The effects of line spacing on objective legibility were insignificant, but participants indicated that their subjective legibility was impaired at extreme values ($0.8$ and $1.8$). The authors summarized that increasing the font size is an efficient way to improve legibility.

\subsection{Performance Optimization }
\label{sec:performanceOptimization}
Improving a web application's energy efficiency, related to many other performance metrics, requires a deep understanding of various optimization techniques. Therefore, for instance, Wagner \cite{wagnerWebPerformanceAction2017} collated several performance improvement techniques that were also addressed in research. Such as optimizing CSS as well as JS content (e.g., \cite{caoSolutionWebFontend2017}), tackling the problem of media-related optimization (e.g., \cite{willisLowPowerWeb2020a}), considered different transmission protocols (e.g., \cite{guptaSurveyApplicationLayer2021}), covering design-effective aspects (e.g., \cite{liCrosslayerOptimizationVideo2023})  and also mentioned the optimization of web fonts considering mainly the application layer of the \ac{OSI}. Riet et al. \cite{vanrietOptimizeWayIndustrial2023} further advanced these techniques by incorporating some of them into a replicable performance engineering plan consisting of 13 interventions for the desktop and mobile web. Appropriating those to a sample case study showed significant performance improvement opportunities. One of those was "Intervention 10: Preload Fonts", also known as lazy-loading. Besides the papers mentioned above, we also reviewed some gray literature that described other font-related techniques, as follows:
\vspace{-0.4cm}
\subsubsection{Font Subsetting: } It is a technique for reducing the size of a font file. Here, only the characters needed in a font file are selected, and the rest is discarded. One often-used example is subsetting a font by language, e.g., to provide a font with only Latin characters for English-language pages. Using this technique, the loading time of the fonts can be improved by more than 200\% \cite{wagnerWebPerformanceAction2017}. 
The widely used Google Web Fonts API is able to automatically create a subset for many font families by providing an additional attribute. There is also the possibility to create a subset specifically for custom purposes by modifying the file \cite{barashkov2022}. Last but not least, the CSS \textit{unicode-range} property of a \texttt{@font-face} Definition specifies the characters for which, if any, the font is to be loaded \cite{wagnerWebPerformanceAction2017}. 
\vspace{-0.4cm}
\subsubsection{Font Hosting: } There are two ways to load fonts: Self-Hosting or Third-Party Hosting. While self-hosting stores the files on your own web server, third-party hosting uses a font service such as Google Fonts \cite{morey2022}. While using Third-Party-Hosting is generally considered easier, it requires additional communication with an external resource, resulting in a decreased loading speed and a dependency on the service provider \cite{liew2021}.
\vspace{-0.4cm}
\subsubsection{Font Loading: }Another option is using \texttt{$<$link rel="preload"$>$} so that the font is not loaded when it is encountered in the external stylesheet but already when this tag appears. Another variant is to use the Font Loading API in the JavaScript code. This way, the process can be followed precisely, and user-defined steps can be initiated \cite{grigorik2020}.
\vspace{-0.4cm}
\subsubsection{Font Rendering:} The \textit{font-display} CSS attribute determines what happens until the external font file is loaded: Should the browser wait until it is loaded or render the text in a fallback font? Besides these two options, there are others to choose from. The default behavior varies from browser to browser. The attribute affects the \textit{Largest Contentful Paint}, the \textit{First Contentful Paint}, and the stability of the layout \cite{webdev}.
\vspace{-0.4cm}
\subsubsection{General Optimizations:} Furthermore, the use of general optimization strategies is often suggested, such as enabling client-side caching \cite{grigorik2020} or enabling server-side compression using algorithms such as GZIP or Brotli. The latter should only be used for TTF and OTF formats, as WOFF and WOFF2 already use built-in compression \cite{kaleev2023}.

Many studies have been conducted, focusing on several aspects of performance improvement or improving the visual perception of fonts. However, we have not identified any scientific literature yet examining the performance of font formats in terms of several efficiency criteria. Also, the grey literature in this field has no benchmarking tests available. The present study attempts to fill this gap.

\section{Experimental Methodology}
\label{sec:experimentalMethodology}
\begin{figure}[!ht]
    \centering
    \includegraphics[height=0.5\textwidth]{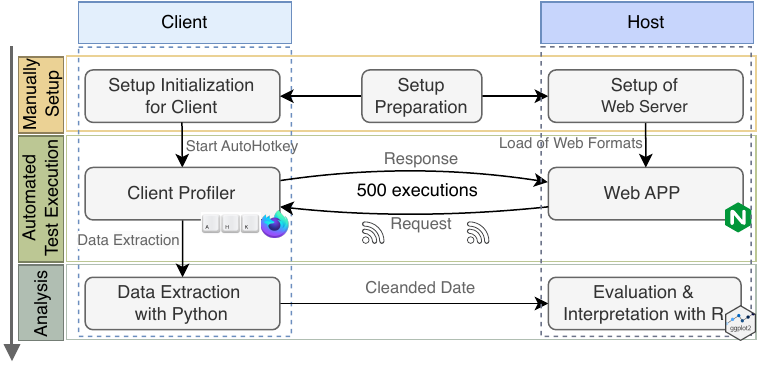}
    \caption{Overview of experimental setup.}
    \label{fig:methodology}
\end{figure}

To assess the influence of various font formats on the performance of web applications and address the research questions as well as the hypothesis (see Section \ref{sec:introduction}), we have outlined the following proposed experimental framework, simplified in Fig. \ref{fig:methodology}.
\subsection{Environmental Setup}
The web client (Windows 11; AMD Ryzen 5 5500, 32GB) and web server (Ubuntu 22; i7-6700HQ, 16GB) are hosted on two different machines inside the same \ac{WLAN}. The web server uses NGINX to handle the requests and deploy the web application containing the different web formats.

The web application consists of a single web page containing several headings together with textual paragraphs and a footer. In total, three different fonts are used as this is a common practice for designing interfaces \cite{tidwellDesigningInterfacesPatterns2020}. The following combination, based on a selection of common Google Fonts, was selected:
\begin{itemize}
    \item \textit{Raleway Extrabold v3.000} for the headings, 
    \item \textit{SourceSans Regular v3.052} for the paragraphs, and
    \item \textit{Montserrat Semibold v3.100} for the footer.
\end{itemize}

Each of the fonts is provided in each of the considered font formats: \textit{\ac{TTF}}, \textit{\ac{OTF}},\textit{\ac{WOFF}}, and \textit{\ac{WOFF2}}. While all of the required formats of Montserrat are downloaded directly from the official Github repository \cite{ulanovsky2021}, the repositories of Raleway \cite{mcInerney2016} and SourceSans \cite{hunt2023} did only provide \ac{TTF} and \ac{OTF} versions. For obtaining the \ac{WOFF} and \ac{WOFF2} formats, which contain the same information as their counterparts, two prevalent NPM conversion tools \textit{ttf2woff version 3.0.0} \cite{ttf2woff} and \textit{ttf2woff2 version 5.0.0} \cite{ttf2woff2} are used to convert from TTF to WOFF as well to WOFF2. Both packages claim over $150K$ weekly installations.

Performance measurements are subject to fluctuations on the client device due to other processes running on the same client machine. These processes have different resource consumption at different times. To mitigate this specific threat, some common practices \cite{dornauerEnergySavingStrategiesMobile2023c} were considered. First, the client device is restarted before each bunch of repeated trials to minimize the influence of processes running and ensure the trials are independent. After the reboot, all heavy background processes are shut down, including collaboration tools, synchronization software, antivirus software that might perform arbitrary security scans, and background browser processes. These measures are intended to put the computer into a "low idle energy fluctuations" state before profiling is started.

\subsection{Automated Test Execution}
A total of 500 individual trials were conducted for each font format, wherein the resource consumption was meticulously profiled. To measure the performance of the web application, a commonly used tool was applied, the \textit{Firefox Profiler}\cite{firerfoxprofiler}. Specifically, the developmental version \textit{Firefox Nightly} (version 115.0a1) is used, as memory usage analysis is only available in the respective Firefox Profiler. In this way, we were able to analyze and measure various performance metrics of the entire browser process or specific threads. It provides insights into CPU, memory, as well as energy consumption. This study did not consider other browsers, as they lack support for the aforementioned performance metrics. 

The selected profiler takes samples at a desired interval and writes the result to a buffer, overwriting old values when complete. The sampling rate in this experiment was set to 0.5 milliseconds, with a buffer size of 2 GiB. In addition, the following manual settings are applied to all experimental runs in order the retrieve the specific metrics: any additional threads have been disabled, Browser Cache deactivated, only CPU Utilization has been enabled in the "Features" section, and the experimental features "Process CPU Utilization" and "energy Use" have been switched on. 

The repeated trials are conducted automatically using AutoHotkey \cite{grayAutoHotkey}. The respective script launches Firefox Nightly, starts the Firefox Profiler, navigates to the web application, stops the Profiler, saves the result as JSON, and shuts down Firefox before the next run starts.

Some actions have been performed to minimize skewed results in the runs. Causes of such distortions include, for example, the Profiler recording the resource consumption of other processes and tasks or the runs influencing each other. For this reason, the following delays are built into the AutoHotkey script: an 8-second delay after Firefox Nightly is started and an 8-second delay after it is closed.

Each batch of runs, where each font is provided with the same format, is automatically profiled 500 times, and the results are used for evaluation.

\subsection{Data Evaluation Process}
The data provided by one run of the Firefox Profiler includes performance metrics, events, and other actions that happened during the recorded time frame. The Profiling was started before the request and stopped after it finished. While each metric covers the entire recorded period, only a fraction of that period is necessary for examining performance. To focus on the font acquisition process, including decoding, conversion, and display, the data series of each performance metric is trimmed accordingly. The starting point is determined by extracting the time in milliseconds when the "DOMContentLoaded" event occurs. This event indicates that the HTML file has been downloaded, parsed, and external resources are being fetched \cite{DOMContentLoaded}. Similarly, the end time is determined by the millisecond timestamp when the \textit{Load Event} is triggered. At this point, all resources, including fonts, have been successfully loaded and rendered\cite{load}. 

The profile of one run provides many measurement values, specifically yielding information about the performance. Some of these metrics are expressed as related to a particular thread. This allows a targeted analysis of that specific thread in the browser process, which particularly processes the request to the test web application, and further excludes any values of other threads. However, other measurements, such as the energy consumption are provided per core and in total. When evaluating the performance, the  metrics given in Table \ref{tab:performanceMetrics} are extracted using a Python script and considered in the evaluation. 

Although a sampling rate of 0.5 milliseconds was set, the distances between the individual samples are not always uniform. For this reason, the values are first converted to \textit{per milliseconds} to allow for comparability. Subsequently, for each data series provided, only the values that lie within the desired observation period, i.e., between the two events, are considered. 

\setlength{\tabcolsep}{0.1cm}
\begin{table*}[htpb]
\centering
\caption{Performance-related metrics overview.}
\label{tab:performanceMetrics}
\begin{tabular}{p{1.5cm} p{8.5cm} p{1.6cm}}
\hline
\rowcolor[rgb]{0.753,0.753,0.753} 
\multicolumn{1}{c}{Metric} & \multicolumn{1}{c}{Description} & \multicolumn{1}{c}{Unit}\\
\hline
Load Time &    
The time in Milliseconds between the events \textit{DOMContentLoaded} and \textit{Load}. &
Milliseconds \textbf{[ms]} \\
\hline
CPU Cycles &
The Profiler`s JSON output provides a data series for each thread, indicating the number of processor cycles required between each sampling point. For each run, the sum of this data series between the two events considered is extracted. & 
CPU Cycles \textbf{[Count]} \\
\hline
Average memory allocation changes & 
This metric provides the number of relative changes in the allocated memory. The size of the allocated memory at a given sampling point in time can be calculated by the cumulative sum of the changes. However, since an investigation of the performance is primarily concerned with the development of memory usage, absolute values are not required and the 10\% trimmed mean of the series of relative values is sufficient for analysis. & 
MegaByte \textbf{[MB]}\\
\hline
Energy Consumption & 
The energy consumption provided by an implementation of RAPL (Running Average Power Limit) is used \cite{hackenbergPowerMeasurementTechniques2013} and is specified in picowatt-hours by the Firefox Profiler. The sum of all data series values within the two events is considered. & 
Milli-Watt-hours \textbf{[mWh]}\\
\hline
\end{tabular}
\end{table*}
\newpage 
Following this, a one-way ANOVA was used, along with the Tukey HSD post-hoc test, to assess the significance of differences in font formats for various hardware-related metrics. To meet ANOVA preconditions, both the Brown-Forsythe test for variance equality \cite{brownRobustTestsEquality1974} and the Lilliefors-Test for normality testing \cite{lillieforsKolmogorovSmirnovTestNormality1967} (with transformation using outlier exclusion by Tukey's Outer Fence and Box-Cox-Transformation mentioned by \cite{johansenSimpleTransformationIndependent2018}) were conducted, combined with graphical interpretation. According to Blanca et al., the one-way ANOVA is relatively robust against violations of the normality assumption. Therefore, the transformation mentioned seems justifiable. For details, we point out the replication package published on Zenodo \cite{dornauerRoleFontFormats2023}. 
\section{Results}
\label{sec:results}
The forthcoming analysis provides a comprehensive evaluation of hardware resource consumption, encompassing load times, processor cycles, memory allocations, and energy consumption. For each metric, the Brown-Forsythe test confirmed equal group variances among font formats, ensuring the validity of the homogeneity of variance assumption for one-way ANOVA. Additionally, the Kolmogorov-Smirnov-Lilliefors test verified the normality of the data.

\begin{figure}
  \begin{subfigure}{1\textwidth}
      \centering
      \begin{tabular}{p{2cm} r r} 
    \hline
    \rowcolor[HTML]{C0C0C0} \multicolumn{1}{c}{Format} & \multicolumn{1}{c}{Load Time [ms]} & \multicolumn{1}{c}{Processor Cycles [Count]}\\ 
    \hline
    {\cellcolor[HTML]{a8b1bd}}otf & 196.0 (189.1, 205.1) & 32667944 (31923896, 33319052) \\
    {\cellcolor[HTML]{faab6b}}ttf & 210.0 (203.5, 219.1) & 32217562 (31500902, 32846721) \\
    {\cellcolor[HTML]{fad46b}}woff & 172.0 (164.9, 180.1) & 28661862 (28039596, 29193426) \\
    {\cellcolor[HTML]{e1d0b7}}woff2 & 161.5 (155.8, 169.3) & 27888552 (27333156, 28507697) \\ 
    \hline 
    \rowcolor[HTML]{C0C0C0} \multicolumn{1}{c}{Format} & \multicolumn{1}{c}{Memory Allocations [Mb]} & \multicolumn{1}{c}{Energy Consumption [mWh]}\\ 
    \hline
    {\cellcolor[HTML]{a8b1bd}}otf & 2264.5 (1953.7, 2634.3) & 3.13 (2.99, 3.26) \\
    {\cellcolor[HTML]{faab6b}}ttf & 2727.5 (2306.4, 3152.3) & 3.27 (3.15, 3.42) \\
    {\cellcolor[HTML]{fad46b}}woff & 6235.8 (5157.0, 7540.5) & 2.72 (2.57, 2.88) \\ 
    {\cellcolor[HTML]{e1d0b7}}woff2 & 5746.6 (4877.8, 6779.1) & 2.59 (2.47, 2.73) \\
    \hline 
    \end{tabular}
      \caption{Values given as: median($1^{st}$ quartile, $3^{rd}$ quartile)}
      \label{fig:comparison_median_quartile}
    \end{subfigure}

    \begin{subfigure}{.5\textwidth}
      \centering
      \includegraphics[width=\linewidth]{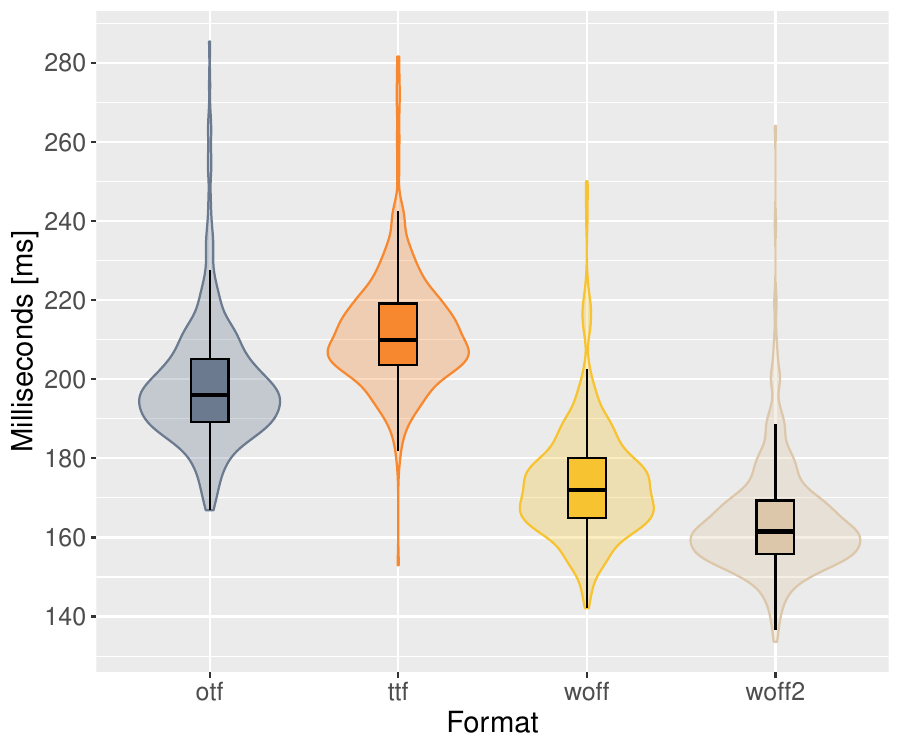}
      \caption{Load Time}
      \label{fig:violin_length_constellation}
    \end{subfigure}
    \begin{subfigure}{.5\textwidth}
      \centering
      \includegraphics[width=\linewidth]{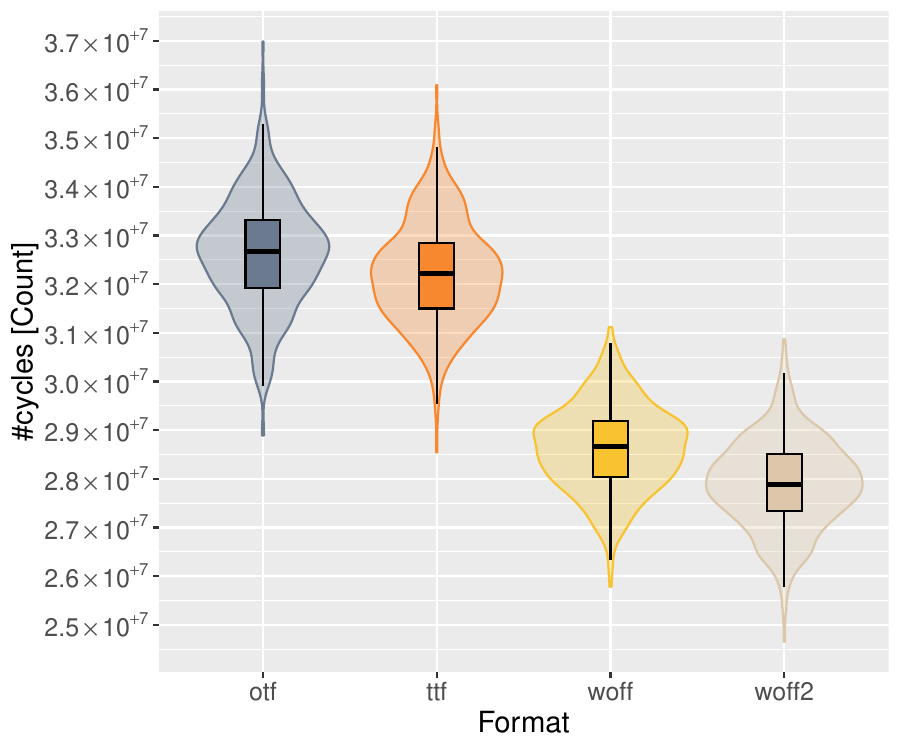}
      \caption{Processor Cycles}
      \label{fig:violin_cycles_constellation}
    \end{subfigure}
     \begin{subfigure}{.5\textwidth}
      \centering
      \includegraphics[width=\linewidth]{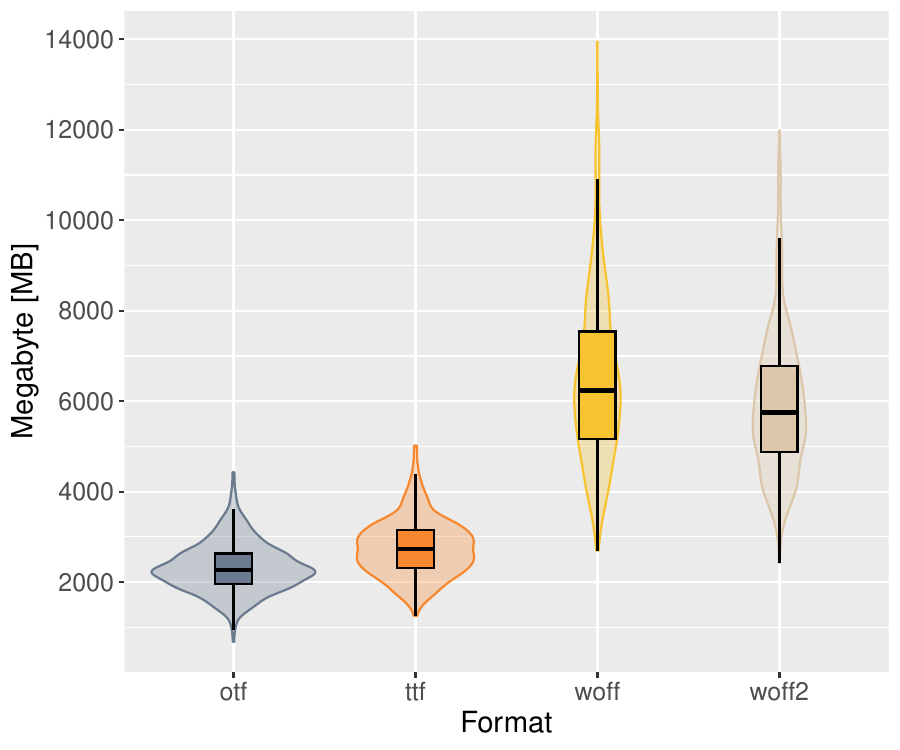}
      \caption{Average memory allocation changes}
      \label{fig:violin_mem_constellation}
    \end{subfigure}%
     \begin{subfigure}{.5\textwidth}
      \centering
      \includegraphics[width=\linewidth]{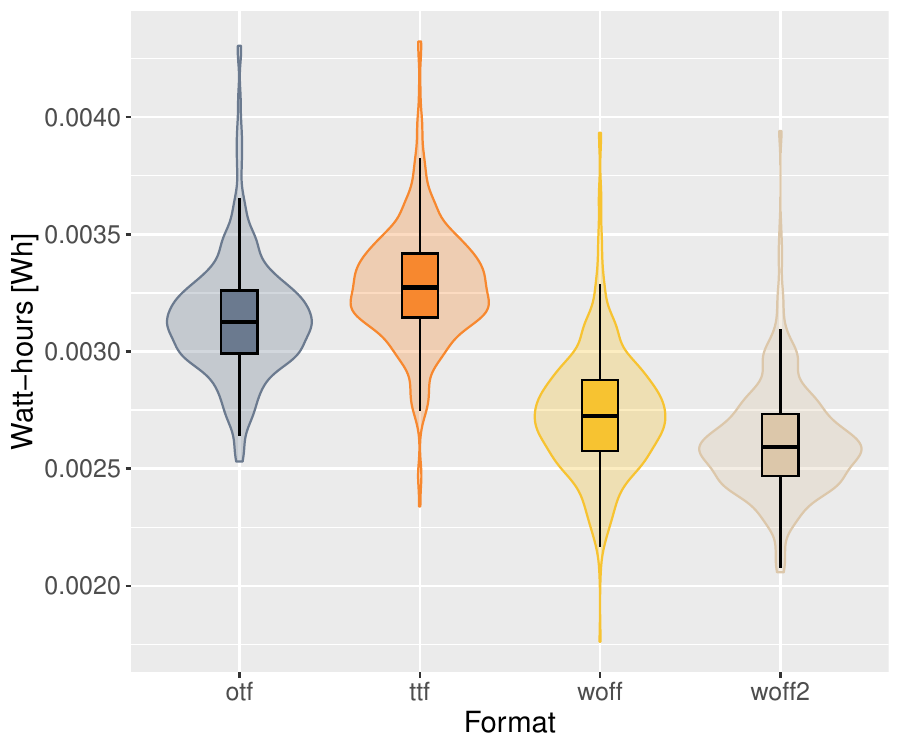}
      \caption{Energy Consumption}
      \label{fig:violin_pow_constellation}
    \end{subfigure}%
\caption{Comparison of performance metrics (b)-(d) across the four \textbf{web font formats}.}
\label{fig:fig}
\end{figure}
\newpage
\subsection{Load Times}
First, a look at the differences in the length of the time frame between each of the font formats is provided. The length of this specifies how much time has been spent after the HTML has been parsed to download further, parse, and render external resources \cite{DOMContentLoaded}. 

As illustrated in Fig. \ref{fig:violin_length_constellation}, there are clear differences in the individual font formats considering the web page speed. TrueType Fonts had the lowest performance in terms of the speed at which the fonts are loaded, parsed, and rendered, with a median of 209  ms. OpenType fonts already had a slightly shorter length between the two window events considered, with a median of 195 ms.  Fonts provided in the Web Open Font Formats were significantly faster in the analysis, with version one yielding a median of 171 ms and its extension WOFF2 showing the best performance with a median of 161 ms. 

Also evident in Figures \ref{fig:violin_length_constellation} is the skewness of the distributions. The skewness of each format is positive, indicating a right-skewed distribution ranging from TrueType with a skewness of 1.11 to WOFF2 with a value of 1.95. These values suggest that more values on the upper end show a longer loading time than the median. Similarly, all four distributions show a positive kurtosis, ranging from 7.00 (TrueType) to 11.2 (WOFF), indicating leptokurtic distributions. This signalizes the required time between both events to contain more extreme outliers than a normal distribution, especially at the upper end. 

The results of the one-way ANOVA revealed that there was a statistically significant difference in the document loading time between at least two formats ($F(3, 1996) = 1138, p < 2.2e^{-16}$). Thus, the null hypothesis $H_0$ can be rejected. A Tukey HSD test shows a significant difference ($p_{adj}=0$) between all pairwise format comparisons. With $\eta^2$ value (and partial $\eta^2$ value) equals 0.631, the effect size of format on load time is considered as high \cite{cohen1988}.

\subsection{Processor Cycles}
As a metric of resource consumption, the results of the processor cycles are presented. An overview of the sum of required processor cycles is shown in Fig. \ref{fig:violin_cycles_constellation}. As illustrated, the variations depend on the font format. In terms of the required processor cycles, using WOFF2 results in the lowest CPU consumption, with a mean of 27.89 million cycles. The second-best performance was achieved by using the WOFF font format, with a median of 28.66 million cycles. This is followed by TrueType fonts (32.21 M) and OpenType fonts (32.67 M). With a correlation coefficient of 0.736, the correlation between the length of the time frame and the required processor cycles is high. 

Considering the skewness of the processor cycles concerning the format, very symmetrical distributions are observed, as the values range from -0.064 (WOFF) to 0.177 (TrueType). Furthermore, the kurtosis of the distributions is similar to those of a normal distribution, as all values are within the range of 2.97 (WOFF) and 3.45 (OpenType).   

The null hypothesis associated with $H_2$ can be rejected, stating that the format does not influence the processor utilization measured in the number of cycles required. The results revealed that there is a statistically significant difference in the processor cycles between at least two formats ($F(3, 1996) = 2903, p < 2.2e^{-16}$). The Tukey HSD test indicates that there are significant differences between all pairs ($p_{adj}=0 < 0.01$). According to the calculated $\eta^2$ value of 0.813, the effect size of the format on the processor cycles required is considered high \cite{cohen1988}. 

\subsection{Memory Allocation}
The profiler results regarding the required memory allocations are presented. An overview of the relative changes to the allocated memory is shown in Fig. \ref{fig:violin_mem_constellation}. It is noted that the values provided are the 10\% trimmed means of the relative memory allocation changes. The distribution of the metric differs depending on the format. While OpenType fonts and TrueType fonts show few additional memory allocations with medians of 2264 and 2728 allocated bytes, respectively, the web fonts have significantly more memory allocations with medians of 6236 (WOFF) and 5747 (WOFF2) allocated bytes. 

The allocation of memory is symmetric for all formats with skewnesses in the range of 0.493 (TrueType) to 0.874 (WOFF2). Similarly, the kurtosis of the distributions is similar to those of a normal distribution, ranging from 3.43 (TrueType) to 4.19 (WOFF2).  

Furthermore, a clear difference in the spread between system fonts and web fonts is illustrated. While system fonts have a standard deviation of 545 (OpenType) and 630 (TrueType) bytes, a greater dispersion is shown for WOFF and WOFF2, whose average deviation from the mean varies by 1846 and 1540, respectively. 

Again, a one-way ANOVA test is performed. The $H_3$ associated null hypothesis states that the format has no effect on the memory allocations required. The results revealed that there was a statistically significant difference in memory consumption between at least two formats ($F(3, 1996) = 1381, p < 2.2e^{-16}$). Further analysis with Tukey HSD shows a significant difference between all pairwise groups. Similarly to the prior resource metrics, the effect sizes are considered as high according to an $\eta^2$ value of 0.678. 

\subsection{Energy Consumption}
Finally, the energy consumption of each of the formats is analyzed by specifying the watts per hour consumed in the considered time frame, i.e., between the two events. A violin plot in Fig. \ref{fig:violin_pow_constellation} is provided.

By only looking at the required watts per hour, web fonts can show a better performance related to energy consumption. WOFF2 formats required the least energy with a median of 2.59 Milliwatts per hour (mWh), followed by the older version WOFF with 2.72 mWh and OTF with 3.13 mWh. The lowest performance in terms of energy consumption was achieved by using TrueType fonts, with a median of 3.27 mWh.  

To test $H_4$, whether the font format affects the energy consumption while the document is loaded, a one-way ANOVA is conducted. The results of this test revealed that there was a statistically significant difference in energy consumption between at least two formats $(F(3, 812) = 1381.  p < 2.2e^{-16})$. The Tukey HSD test shows a significant difference between all pairwise font formats. The effect size is strong according to the $\eta^2$ value of 0.55. Therefore, the associated null hypothesis is rejected, stating that the font format does not influence energy consumption. 
\section{Discussion}
\label{sec:discussion}
Fonts play a crucial role in design and typography. In the competitive market of web applications, they need to have a distinctive appeal and value proposition. Consequently, it has become customary to employ custom-designed fonts in one of the four prevalent formats: TTF, OTF, WOFF, and WOFF2. However, incorporating these font formats can introduce performance challenges that potentially compromise user satisfaction, particularly in regions with limited data bandwidth, such as developing countries. Thus, the selection of an appropriate format assumes paramount importance. Through the analysis, we can reject the null hypotheses associated with $H_1$ to $H_4$, as compelling evidence has emerged highlighting a significant impact of the font format on all four dependent variables.
\vspace{-0.4cm}
\subsubsection{H1:} The results confirm the recommendations mentioned by the gray literature \cite{webdev, stein2022WebAlmanac2022} about the usage of the font formats. The web fonts are able to show a faster loading time in contrast to the system fonts. Usage of the newest Web Open Font Format version 2 resulted in the best loading time, followed by WOFF, OpenType and TTF. 
\vspace{-0.4cm}
\subsubsection{H2:} The process utilization results were contrary to the expectations since a higher CPU utilization was expected for web fonts due to the necessary decompression. The possible reason for these contradictory results is assumed to be the longer network connection, which leads to a longer observation period for system formats with respect to the processor cycles (idle time). The longer network connection may require additional processing cycles, compensating for decompression's necessary disadvantage.
\vspace{-0.4cm}
\subsubsection{H3:} An examination of the performance of the different formats concerning the allocated memory resulted in the conclusion that system fonts (TrueType and OpenType) require significantly less memory than web fonts. Although the observation period is longer for system fonts, OpenType fonts were able to minimize the required allocated memory, with TrueType fonts following closely with 1.2 times more allocations. Using WOFF2 resulted in 2.53 times more memory allocations on average; for WOFF the factor increases to 2.75. To prioritize memory consumption, it is advisable to specify fonts with fallback formats in this order.
\vspace{-0.4cm}
\subsubsection{H4:} The last performance metric examined is the energy consumption required to load the web page with respect to the font formats used. The results show that the usage of font formats with fallback formats in the following order can reduce energy consumption: WOFF2, WOFF, OpenType, and TrueType. Web developers with a high prioritization of energy efficiency are recommended to use this order.

To sum up, we suggest the following key takeaways to both researchers and practitioners: 
\begin{tcolorbox}[breakable,  boxsep = 0pt, outer arc = 0pt]
\begin{enumerate}
\item TTF and OTF fonts have been established for a considerable period, making them widely adopted and offering an extensive range of styles.

\item Overall, WOFF2 is the favored font format due to its superior performance, as predicted by gray literature (e.g., \cite{webdev, stein2022WebAlmanac2022,webdev}). Hence, a wrapper like \texttt{ttf2woff2} seems to be appropriate and recommended.

\item It must be noted that WOFF2 might face limited support in specific browsers. As a result, it is advisable to consider using WOFF as a fallback option to ensure broader compatibility.
\end{enumerate}
\end{tcolorbox}

\section{Threats to Validity}
\label{sec:threatsToValididty}
Threats to Validity refer to factors that may undermine the reliability and generalizability of this study. We split those into four types based on Cook and Campbell  \cite{cookQuasiexperimentationDesignAnalysis1979} that we want to discuss further: 
\vspace{-0.4cm}
\subsubsection{Internal Validity:} In order to mitigate this risk, we took into consideration the requirements for reliable benchmarking from Beyer et al. \cite{beyerReliableBenchmarkingRequirements2019}. Concretely this means we decided to completely automate the test execution using AutoHotkey, which also allows independent replication and verification of the experiment. Similarly, we have used common practices in order to keep the performance measurement stable using, for instance, a timeout window between the runs, a specific configuration to avoid external influences, and conducted 500 test runs per web font format to recognize potential outliers.
\vspace{-0.4cm}
\subsubsection{External Validity: } As there exist many font styles, we combined three font styles as this is a typical design principle 
\cite{tidwellDesigningInterfacesPatterns2020}. Nonetheless, while our combination encompasses a variety of font styles, it is worth considering the inclusion of further styles as a potential avenue for future research. Moreover, to retrieve the targeted performance metrics, our study was constrained to a concrete experimental setup (described in Section \ref{sec:experimentalMethodology}). This limitation adversely impacts the generalizability of our findings, given the multitude of alternative browsers and client devices available in the market. However, it is noteworthy that our results confirm prevailing assumptions documented in the grey literature, lending support to their relevance across different experimental configurations. 
\vspace{-0.4cm}
\subsubsection{Construct Validity: } In terms of the construct design and poor operationalization, we outlined the design process. We selected the performance metric in a way that direct impacts to hardware effects were identifiable. Specifically, CPU cycles, allocated and deallocated memory, and the loading time. For a general metric, we selected the energy consumption as an overall is expected to provide a good overview of how efficient the website is in providing the expected outcome. 
\vspace{-0.4cm}
\subsubsection{Statistical Conclusion Validity:} This threat is addressed by using statistical hypothesis tests, namely \textit{one-way ANOVA} and \textit{Tukey HSD}, such that the conclusions drawn from this study are founded on common data analysis practices. Furthermore, the results are checked against grey literature, which was selected by specific inclusion criteria. Furthermore, we have fully disclosed all findings and test materials in Zenodo  \cite{dornauerRoleFontFormats2023} 
obtained by our experiments. 

\section{Conclusion and Future Work}
\label{sec:conclusion}
Loaded via CSS, font formats give web developers the opportunity to individualize their visual representation, with the drawback of performance downturns. Hence, the selection of an appropriate font format is essential. This study aims to provide insights into the prevailing font formats, \ac{TTF}, \ac{OTF}, \ac{WOFF}, as well as its second generation. Our benchmarking shows that \ac{WOFF2} surpasses all other types, albeit with a higher memory allocation. Thus, we conclude that practitioners should employ \ac{WOFF2} and consider converting other formats to \ac{WOFF2} when feasible. Other optimization options mentioned in chapter \ref{sec:relatedWork}, such as font subsetting or the hosting method used, also influence the web application's performance and could be part of further elaboration.
\section{Acknowledgment}
This work has been supported by and done in the scope of the ITEA3-SmartDelta project, which has been funded by the Austrian Research Promotion Agency (FFG, Grant No. 890417). 
\bibliographystyle{splncs04}
\bibliography{bibliography}
\end{document}